\documentclass[12pt,a4paper]{article}
\usepackage[cp1250]{inputenc}
\usepackage[T1]{fontenc}
\usepackage{amsmath}
\usepackage{amsfonts}
\usepackage{amssymb}
\usepackage{amsthm}
\usepackage{graphicx}
\usepackage{rotating}
\usepackage{multirow}
\usepackage{multicol}
\usepackage{tabularx}
\usepackage{color}
\usepackage{overcite}
\usepackage[ruled,linesnumbered]{algorithm2e}
\graphicspath{ {./figures/} }   
\usepackage{subcaption}
\usepackage{color}   
\usepackage[hyperindex]{hyperref}
\usepackage{verbatim}
\usepackage{mdsymbol}
\usepackage{breakcites}   

\hypersetup{colorlinks=true, citecolor=blue, linkcolor=red}
\usepackage{natbib}

\usepackage[left=1in,top=1in,right=1in,bottom=1in,nohead,paperwidth=8.5in, paperheight=11in]{geometry}

\theoremstyle{definition}

\title{\bf Sparse twoblock dimension reduction for simultaneous compression and variable selection in two blocks of variables}
	\vspace{3mm}
	\author{Sven Serneels$^{1,2}$}

\begin{document}
\maketitle
\noindent
{\small
$^{1}$Aspen Technology, Bedford, MA, USA\\
$^{2}$Department of Mathematics, University of Antwerp, Antwerp, Belgium\\

\begin{abstract} \label{Abs}
A method is introduced to perform simultaneous sparse dimension reduction on two blocks of variables. Beyond dimension reduction, it also yields an estimator for multivariate regression with the capability to intrinsically deselect uninformative variables in both independent and dependent blocks. An algorithm is provided that leads to a straightforward implementation of the method. The benefits of simultaneous sparse dimension reduction are shown to carry through to enhanced capability to predict a set of multivariate dependent variables jointly. Both in a simulation study and in two chemometric applications, the new method outperforms its dense counterpart, as well as multivariate partial least squares.     
\bigskip

\noindent \textbf{Keywords: }{PLS, XY-PLS, Sparse Twoblock, Prediction, Dimension Reduction}
    
\end{abstract}

\newpage

\section{Introduction}
\label{sec:Int}
         
In an era of increasing data dimensions, it becomes ever more important to be able to summarize information in data into just a few dimensions. This holds true, regardless if data consist of one single block, or of multiple blocks. In the latter context, it can be important to simultaneously reduce the dimensions of two or more blocks in such a way that information on the interplay between the blocks of variables is summarized into a lower dimensional space. This becomes most important when combined with a regression task, when the objective is to predict one block from the other. 

The objective to identify a lower dimensional space as a set of linear combinations of the original variables that contain all information relevant to model one or more dependent variables, is known as {\em sufficient dimension reduction} (SDR). Let $\mathbf{X}\in\mathbb{R}^{n\times p}$ be a data matrix that consists of $n$ observations of a $p$ variate independent random variable $\mathbf{x}$, and likewise, let $\mathbf{Y}\in\mathbb{R}^{n\times q}$ denote the matrix of $n$ observations of a $q$ variate dependent random variable $\mathbf{y}$. Then sufficient dimension reduction aims to estimate latent variables $\mathbf{T} = \mathbf{X}\mathbf{W}$ as linear combinations of the original variables that satisfy: 
\begin{equation}\label{equation:SDR}
\mathbf{Y} \upvDash \mathbf{X} \ | \ \mathbf{T},
\end{equation}
where $\upvDash$ denotes statistical independence.
The subspace that fulfils the above equation, is called the {\em central subspace}. Many estimators of the central subspace have been proposed, but one of the oldest and still most computationally efficient ones, is partial least squares (PLS). PLS had existed decades before most publications on SDR, but the proof that PLS scores present an estimate of the central subspace, was only presented about a decade ago by \cite{cook2013envelopes}. As the independence of the dependent variable and the central subspace's complement holds in absence of linearity requirements, it should not come as a surprise that PLS dimension reduction is serviceable in the presence of nonlinearity \citep{cook2021pls}.   

Equation \eqref{equation:SDR} defines sufficient dimension reduction, when the objective is to reduce {\em one} block of variables into a smaller space relevant to a dependent variable whose dimensionality need not be reduced. An equally interesting objective, though, is to look at dimension reduction in both independent and dependent blocks {\em simultaneously}, such that the reduced spaces contain all information sufficient to model one another mutually, yet are independent of the complement in either of these spaces. In brief, this implies to look for linear combinations $\mathbf{U} = \mathbf{Y}\mathbf{V}$ and $\mathbf{T} = \mathbf{X}\mathbf{W}$ in the independent and dependent blocks, respectively, that satisfy: 
 
\begin{subequations}\label{equation:MSDR}
\begin{equation}
\mathbf{Y} \upvDash \mathbf{X} \ | \ \mathbf{T},
\end{equation}
\begin{equation}
\mathbf{X} \upvDash \mathbf{Y} \ | \ \mathbf{U}.
\end{equation}
\end{subequations}

Since independence in \eqref{equation:MSDR} is hard to prove, the condition of independence is typically relaxed to a condition of zero covariance. Let $\mathbf{P} = \mathbf{X}^T\mathbf{T}\left(\mathbf{T}^T\mathbf{T}\right)^{-1}$ and $\mathbf{Q} = \mathbf{Y}^T\mathbf{U}\left(\mathbf{U}^T\mathbf{U}\right)^{-1}$ denote the entities commonly referred to as the ($\mathbf{X}$ and $\mathbf{Y}$) {\em loadings}, then the projection of $\mathbf{X}$ onto its dimension reduced subspace is given by $\mathbf{TP}^T$ and its complement by $\mathbf{E} = \left(\mathbf{I}_n - \mathbf{T}\left(\mathbf{T}^T\mathbf{T}\right)^{-1}\mathbf{T}^T\right)\mathbf{X}$. Likewise, in the $\mathbf{Y}$ space, the projection onto the dimension reduced subspace is given by $\mathbf{UQ}^T$ and its complement by $\mathbf{F} = \left(\mathbf{I}_n - \mathbf{U}\left(\mathbf{U}^T\mathbf{U}\right)^{-1}\mathbf{U}^T\right)\mathbf{Y}$. Then the conditions of linear independence that define simultaneous dimension reduction of both blocks, are given by:     
\begin{subequations}\label{eqs:MSDRcov}
\begin{equation}\label{eq: covind}
\mathop{\mbox{cov}}\left(\mathbf{E},\mathbf{F}\right) = \mathbf{0},
\end{equation}  
\begin{equation}\label{eq: covdep}
\mathop{\mbox{cov}}\left(\mathbf{UQ}^T,\mathbf{F}\right) = \mathbf{0}
\end{equation}  
and
\begin{equation}\label{eq: covindep}
\mathop{\mbox{cov}}\left(\mathbf{TP}^T,\mathbf{E}\right) = \mathbf{0}
\end{equation}  
\end{subequations}     

Multivariate PLS regression (PLS2) may seem to yield estimators that satisfy the conditions in Equations \eqref{eqs:MSDRcov} and many applications can be found in the literature that implicitly surmise that PLS2 performs a simultaneous dimension reduction of both blocks of data. However, \cite{cook2023partial} showed that not to be the case. In fact, entities like the ``$\mathbf{Y}$ scores'' and ``$\mathbf{Y}$ loadings'' are nothing more than a computational trick and do not represent entities that are the result of sufficient dimension reduction (while the corresponding $\mathbf{X}$ block entities are). To remedy this, \cite{cook2023partial} proceed by proposing a two-block {\em XY}-PLS algorithm that does provide estimates of the central subspace of simultaneous sufficient dimension reduction for both blocks, as well as regression coefficients that directly model the relation between both blocks. In a simulation study, they show that multivariate regression based on simultaneous dimension reduction of both blocks outperforms both PLS2 and a set of univariate regressions. 

The results from \cite{cook2023partial} are not widely cited and one motivation for this paper is to draw renewed attention to the XY-PLS algorithm. However, like all PLS algorithms, XY-PLS is an algorithm based on {\em dense} data assumptions, i.e. that all variables contribute {\em some} information to the estimation of subspaces and regression coefficients. In practice, uninformative variables may be present in both blocks of data. When left in the model, they may distort estimates or inflate variance of predictions. Therefore, in this paper, a sparse estimator of XY-PLS is introduced, which allows to perform intrinsic variable selection on either of both blocks, or both, with separate tuneable sparsity parameters for each block. Notably, the necessity to develop a sparse version of the XY-PLS algorithm is already implicitly attested to in the third of the example studies in \cite{cook2023partial}, as they decide to manually select only 69 out of 700 predictor variables. In this article, the newly introduced sparse XY-PLS will be shown to outperform dense XY-PLS when uninformative variables are present, both in simulations and in two real world examples. Moreover, as the new method has tuneable sparsity parameters for both blocks, the optimal subset of informative variables can be determined through cross-validation, which obsolesces the need to select a subset of variables arbitrarily. Finally, this paper will also present a straightforward algorithm for the new method, that can be implemented efficiently. 

\section{Sparse Twoblock Estimators}\label{sec:stwopls}

This Section will recap the XY-PLS algorithm proposed in \cite{cook2023partial} and then introduce the new sparse twoblock estimator. 

\subsection{The XY-PLS algorithm}\label{sec:xypls}

After some manipulations, \cite{cook2023partial} derive an algorithm to calculate an estimator that satisfies Equations \eqref{eqs:MSDRcov}. The gist of the idea to assure zero covariance between successive components in both blocks simultaneously, as well as maximizing covariance between the components in the two blocks, resides in performing deflation separately for each block, thereby allowing the dimension reduced space for the two blocks to be based upon a different number of components. 
  
Let $\mathbf{E}_0 = \mathbf{X}$ and $\mathbf{F}_0 = \mathbf{Y}$ and without loss of generality, assume that these and other data matrices are centred. The XY-PLS weighting vectors are defined for each block according to optimization criteria that involve the cross-covariance matrix $\mathbf{S}_{XY} = \mathbf{X}^T\mathbf{Y}$:   

\begin{subequations}\label{eq:critone}
\begin{equation}\label{eq:critonex}
\begin{aligned}
\mathbf{v}_1=& \quad \mathop{\mbox{argmax}}_{\mathbf{a}} & \left(\mathbf{a}^T\mathbf{Y}^T\mathbf{X}\mathbf{X}^T\mathbf{Y}\mathbf{a}\right)\\
 & \ \quad \textrm{s.t.} \quad & \parallel\mathbf{v}_i\parallel=1,\\
 &  & \mbox{cov}(\mathbf{Y}\mathbf{v}_i,\mathbf{Y}\mathbf{v}_j)=0, j < i.\\
\end{aligned}
\end{equation}
and
\begin{equation}\label{eq:critoney}
\begin{aligned}
\mathbf{w}_1=& \quad \mathop{\mbox{argmax}}_{\mathbf{a}} & \left(\mathbf{a}^T\mathbf{X}^T\mathbf{Y}\mathbf{Y}^T\mathbf{X}\mathbf{a}\right)\\
 & \ \quad \textrm{s.t.} \quad & \parallel\mathbf{w}_i\parallel=1,\\
 &  & \mbox{cov}(\mathbf{X}\mathbf{w}_i,\mathbf{X}\mathbf{w}_j)=0, j < i.\\
\end{aligned}
\end{equation}
\end{subequations}

The above objective can be solved analytically and leads to the result that $\mathbf{v}_1$ and $\mathbf{w}_1$ are the dominant left and right singular vectors of $\mathbf{S}_{XY}$. To arrive at successive components, while respecting the zero covariance constraint for the scores, the original data are deflated, analogous to several PLS algorithms. At first, scores are calculated:    

\begin{subequations}\label{eq:tu}
\begin{equation}\label{eq:u}
\mathbf{u}_j = \mathbf{F}_{j-1}\mathbf{v}_j,
\end{equation}
\begin{equation}\label{eq:t}
\mathbf{t}_i = \mathbf{E}_{i-1}\mathbf{w}_j. 
\end{equation}
\end{subequations}
Loadings are calculated as: 
\begin{subequations}\label{eq:pq}
\begin{equation}\label{eq:q}
\mathbf{q}_j = \frac{\mathbf{F}_{j-1}^T\mathbf{u}_j}{\mathbf{u}_j^T\mathbf{u}_j},
\end{equation}
\begin{equation}\label{eq:p}
\mathbf{p}_i = \frac{\mathbf{E}_{i-1}^T\mathbf{t}_i}{\mathbf{t}_i^T\mathbf{t}_i}.
\end{equation}
\end{subequations} 
Finally, the input matrices are deflated into  
\begin{subequations}\label{eq:EF}
\begin{equation}\label{eq:F}
\mathbf{F}_j = \mathbf{F}_{j-1}-\mathbf{u}_j\mathbf{q}_j^T,
\end{equation} 
\begin{equation}\label{eq:E}
\mathbf{E}_i = \mathbf{E}_{i-1}-\mathbf{t}_i\mathbf{p}_i^T. 
\end{equation}
\end{subequations}
Equations \eqref{eq:EF} guarantee that successive scores will be uncorrelated, thereby satisfying \eqref{eq: covindep} and \eqref{eq: covdep}, as well as the side constraints in \eqref{eq:critone}. 

Finally, \cite{cook2023partial} showed that these entities can be recombined to provide a direct estimate for the regression coefficients relating both blocks of input data to each other: 
\begin{equation}\label{eq:regcoeffs}
\mathbf{B} = \mathbf{W}\left(\mathbf{W}^T\mathbf{X}^T\mathbf{X}\mathbf{W}\right)^{-1}\mathbf{W}^T\mathbf{X}^T\mathbf{Y}\mathbf{V}\mathbf{V}^T.
\end{equation}

\subsection{The Sparse Twoblock PLS Algorithm}\label{sec:stwoplsalgo}

To arrive at a sparse set of weighting vectors, we can proceed by inducing sparsity on each of the estimated $\mathbf{v}_j$ and $\mathbf{w}_i$ analogous to how the weighting vectors are sparsified in the SNIPLS algorithm. SNIPLS was introduced as a building block in the algorithm to compute sparse partial robust M regression \citep{hoffmann2015sparse}, but was only later presented as a stand-alone algorithm in \citep{hoffmann2016sparse}. 
Let $\check{\mathbf{w}} = \mathbf{w}/\parallel\mathbf{w}\parallel$ and $\check{\mathbf{v}} = \mathbf{v}/\parallel\mathbf{v}\parallel$. Then for a given set of sparsity parameters $\eta,\kappa \in [0,1)$, sparse weighting vectors are given by:
\begin{subequations}\label{eq:vwsparse}
\begin{equation}\label{eq:vs}
\tilde{\mathbf{v}}_j = \left(\|\check{\mathbf{v}}_j\| - \kappa\ \mathrm{max}_k \|\check{v}_{jk}\|\right) \odot \mathbb{I}\left(\|\check{\mathbf{v}}_j\| - \kappa\ \mathrm{max}_k \|\check{v}_{jk}\| > 0 \right) \odot \mathrm{sgn}(\check{\mathbf{v}}_j)
\end{equation}
and
\begin{equation}\label{eq:ws}
\tilde{\mathbf{w}}_i = \left(\|\check{\mathbf{w}}_i\| - \eta\ \mathrm{max}_\ell \|\check{w}_{i\ell}\|\right) \odot \mathbb{I}\left(\|\check{\mathbf{w}}_i\| - \eta\ \mathrm{max}_\ell \|\check{w}_{i\ell}\| > 0 \right) \odot \mathrm{sgn}(\check{\mathbf{w}}_i),
\end{equation}
\end{subequations}      
where $\mathbb{I}$ denotes the indicator function that returns 1 if the assessed condition is true and zero otherwise and $\odot$ denotes the Hadamard matrix product. 

Using this sparsity inducing formula, we can now introduce the sparse twoblock PLS algorithm, which is presented as Algorithm 1.

\fbox{\begin{minipage}{0.9\textwidth}
\textbf{ Algorithm 1:} Sparse Twoblock PLS algorithm.\\
\noindent\makebox[\linewidth]{\rule{\textwidth}{0.4pt}}
Initialize the algorithm by $\mathbf{E}_0 = \mathbf{X}$ and $\mathbf{F}_0 = \mathbf{Y}$. Let $\mathbf{l}_1(\cdot)$ denote the function that returns the dominant eigenvector of its argument. Then compute:\\
\noindent\makebox[\linewidth]{\rule{\textwidth}{0.4pt}}
\begin{enumerate}
\item \textsl{Response Reduction}
 \begin{enumerate} 
   \item Select $g \leq \mathrm{min}(\mathrm{rank}(\mathbf{Y}^T\mathbf{Y},n-1))$ and a sparsity parameter $\kappa \in [0,1)$
   \item For $j \in [1,g]$: 
			\begin{enumerate}
			\item $\mathbf{v}_j = \mathbf{l}_1\left(\mathbf{F}_j^T\mathbf{X}\mathbf{X}^T\mathbf{F}_j\right)$
			\item $\check{\mathbf{v}}_j = \mathbf{v}_j / \parallel\mathbf{v}_j\parallel $
			\item $\mathbf{m}_j = \mathbb{I}\left(\|\check{\mathbf{v}}_j\| - \kappa\ \mathrm{max}_k \|\check{v}_{jk}\| > 0 \right)$
			\item $\tilde{\mathbf{v}}_j = \left(\|\check{\mathbf{v}}_j\| - \kappa\ \mathrm{max}_k \|\check{v}_{jk}\|\right) \odot \mathbf{m}_j \odot \mathrm{sgn}(\check{\mathbf{v}}_j)$
			\item $\mathbf{u}_j = \mathbf{F}_{j-1}\tilde{\mathbf{v}}_j,$
			\item $\mathbf{q}_j = \frac{\mathbf{F}_{j-1}^T\mathbf{u}_j}{\mathbf{u}_j^T\mathbf{u}_j} \odot \mathbf{m}_j$
			\item $\mathbf{F}_j = \mathbf{F}_{j-1}-\mathbf{u}_j\mathbf{q}_j^T,$
		\end{enumerate}
	\end{enumerate}
		\item \textsl{Predictor Reduction}
		\begin{enumerate} 
   \item Select $h \leq \mathrm{min}(\mathrm{rank}(\mathbf{X}^T\mathbf{X},n-1))$ and a sparsity parameter $\eta \in [0,1)$
   \item For $i \in [1,h]$: 
			\begin{enumerate}
			\item $\mathbf{w}_i = \mathbf{l}_1\left(\mathbf{E}_j^T\mathbf{Y}\mathbf{Y}^T\mathbf{E}_j\right)$
			\item $\check{\mathbf{w}}_i = \mathbf{w}_i / \parallel\mathbf{w}_i\parallel $
		  \item $\mathbf{n}_i = \mathbb{I}\left(\|\check{\mathbf{w}}_i\| - \eta\ \mathrm{max}_\ell \|\check{w}_{i\ell}\| > 0 \right)$
			\item $\tilde{\mathbf{w}}_i = \left(\|\check{\mathbf{w}}_i\| - \eta\ \mathrm{max}_\ell \|\check{w}_{i\ell}\|\right) \odot \mathbf{n}_i \odot \mathrm{sgn}(\check{\mathbf{w}}_i)$
			\item $\mathbf{t}_i = \mathbf{E}_{i-1}\tilde{\mathbf{w}}_i,$
			\item $\mathbf{p}_i = \frac{\mathbf{E}_{i-1}^T\mathbf{t}_i}{\mathbf{t}_i^T\mathbf{t}_i} \odot \mathbf{n}_i$
			\item $\mathbf{E}_i = \mathbf{E}_{i-1}-\mathbf{t}_i\mathbf{p}_i^T,$
		\end{enumerate}
	\end{enumerate}
	\item Calculate regression coefficients as \newline $\mathbf{B} = \mathbf{W}\left(\mathbf{W}^T\mathbf{X}^T\mathbf{X}\mathbf{W}\right)^{-1}\mathbf{W}^T\mathbf{X}^T\mathbf{Y}\mathbf{V}\mathbf{V}^T.$
		\end{enumerate}
\end{minipage}
}

The right hand multiplication by $\mathbf{m}_j$ or $\mathbf{n}_i$ in the sixth step of the reductions is strictly speaking not necessary to yield sparse weighting vectors and regression coefficients, but it enforces consistence in the nonzero elements of all weighting vectors, loadings and regression coefficients for each latent variable.

Algorithm 1 may appear to have been postulated {\em ad hoc}, yet some theoretical reflections can be made about the nature of the estimator. The estimator being postulated at an algorithmic level above is not unlike PLS, which existed for decades prior to its theoretical underpinnings became more clear. However, as this estimator is still closely related to the PLS family of estimators, some similar theoretical observations can be made. 

The key consideration to bear in mind when looking at theoretical properties, is that the entire estimator derives {\em solely} from measures of covariance. In the case of PLS, this was first described in \cite{serneels2004influence} and later independently confirmed in \cite{cook2013envelopes}. Similar results have been shown for trilinear partial least squares \citep{serneels2005influence} and, more recently, for sparse partial least squares regression \citep{serneels2024elegant}. The approach outlined in these publications could be adopted to also derive a population version of the estimator introduced here, as well as to derive its influence function, which could then be used to estimate variance (of course, in the absence of extended theory, a bootstrap approach could be adopted as well, similar to \cite{serneels2005bootstrap}). However, as the present publication focuses on algorithm and application, such derivations are left to future work. 

\section{Simulation study}\label{sec:simul}

\subsection{Data generation process}

This section will assess how well the estimator combines the tasks of variable selection and prediction in a multivariate response setting. To make that assessment, data are simulated according to a latent variables based data generation scheme. Scores are taken as $n$ samples from a multivariate normal $N(\mathbf{0}_h,\mathbf{I}_h)$, such that the true model complexity (number of latent variables) is known and equal to $h$. Loadings are composed of an informative and an uninformative part: $\mathbf{P} = [\mathbf{p}_0, \cdots, \mathbf{p}_{p_1} \mathbf{0} \cdots \mathbf{p}_2]$,with $p_1$ informative variables sampled from the uniform distribution on [-5,5] and $p_2$ uninformative variables whose loadings equal zero for each of the latent variable dimensions. Input data are then generated as  
\begin{equation} 
\mathbf{X} = \mathbf{T}\mathbf{P}^T + \mathbf{G},
\end{equation}
where $\mathbf{G} \sim N(\mathbf{0}, .01\mathbf{I})$ is an appropriately dimensioned random error term. Similar to loadings, true regression coefficients are composed of an informative and an uninformative part, where the informative part that corresponds to the first $q_1$ dependent variables is sampled from the uniform distribution on [.02,.07] and the remaining $q_2$ variables equal zero exactly. Dependent variables are simulated as
\begin{equation} 
\mathbf{Y} = \mathbf{X}\mathbf{B} + \mathbf{H},
\end{equation}
with $\mathbf{H} \sim N(\mathbf{0}, .01\mathbf{I})$ as well. 

\subsection{Metrics}
To evaluate the quality of the results, the following metrics are considered: 
\begin{itemize}
\item MSEB: 
	\begin{equation}
		\mbox{MSEB} = 1/n \sum\left(\mathbf{B} - \hat{\mathbf{B}}[:,1:p_1]\right)^2, 
	\end{equation}
	i.e. the mean squared error of estimated vs. true regression coefficients that correspond to informative dependent variables; 
	\item FPX: the false positive rate for the predictor block, i.e. the percentage of uninformative variables included in the model that should have been left out; 
	\item FNX: the false negative rate for the predictor block, i.e. the percentage of informative variables left out of the model that should have been included;
	\item FPY: the false positive rate for the predictand block; 
	\item FNY: the false negative rate for the predictand block. 
\end{itemize}

\subsection{Results}

\begin{figure}
\centering
\begin{subfigure}{.4\textwidth} 
\includegraphics[width=\textwidth]{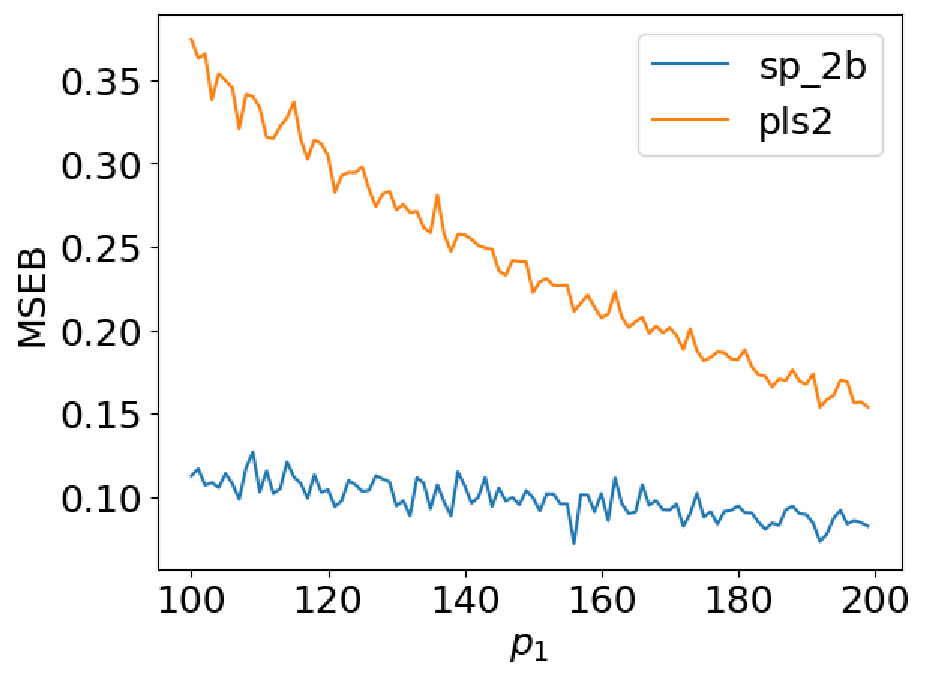}
\caption{}
\label{fig:n_p2_200_MSEB}
\end{subfigure}
\begin{subfigure}{.4\textwidth} 
\includegraphics[width=\textwidth]{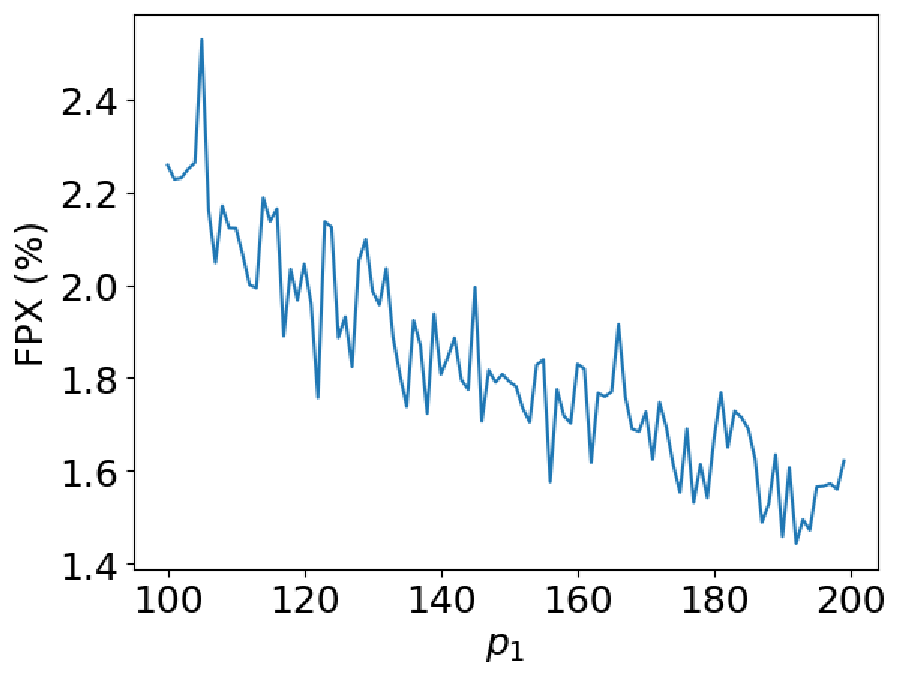}
\caption{}
\label{fig:n_p2_200_FPX}
\end{subfigure}
\begin{subfigure}{.4\textwidth} 
\includegraphics[width=\textwidth]{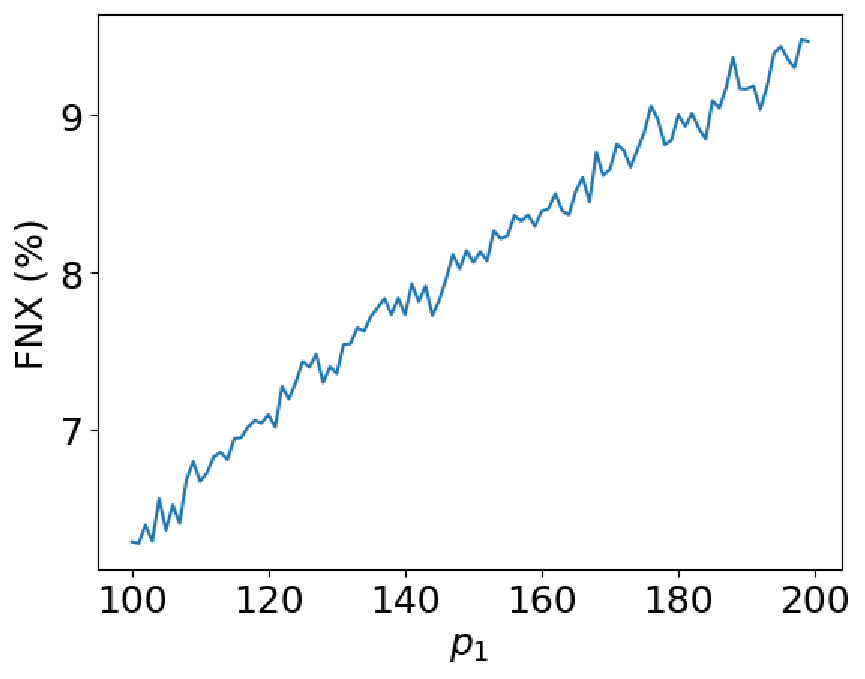}
\caption{}
\label{fig:n_p2_200_FNX}
\end{subfigure}
\begin{subfigure}{.4\textwidth} 
\includegraphics[width=\textwidth]{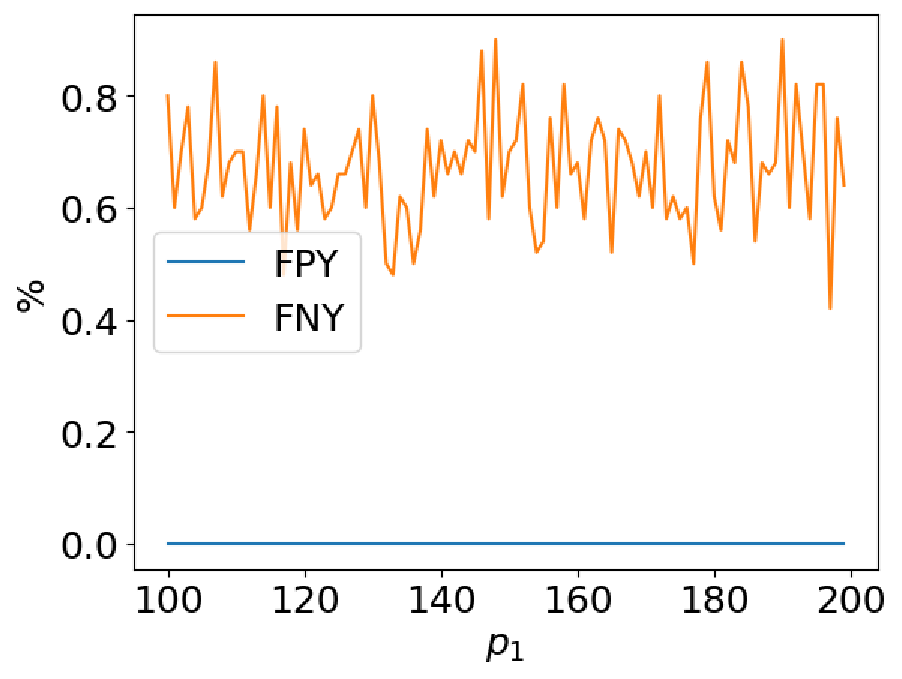}
\caption{}
\label{fig:n_p2_200_FY}
\end{subfigure}
\caption{\label{fig:n_p2_200} Averages over 1000 simulation runs as a function of the number of informative predictor variables $p_1$, of mean squared error of the regression coefficients (a), false positive (b) and false negative (c) rates for the predictor block, as well as false positive and false negative rates for the predictand block (d). In this simulation, $n=100, p_2=200, q_1=3, q_2=2, g=1, h=3$.}
\end{figure}

In what follows, results will be reported as averages of 1000 simulation runs in settings of varying dimensionalities. Results will be reported for three simulation scenarios with different dimensionalities and different proportions of uninformative variables. At first, Figure \ref{fig:n_p2_200} summarizes results for the simulation scenario among the three discussed in which the estimators can be expected to have most challenges.     Figure \ref{fig:n_p2_200} presents the results of a simulation the dimensionality of the predictor block $p_1$ is varied from 100 to 200 informative variables and in each simulation $p_2 = 200$ uninformative variables are added. Each sample consists of $n=100$ cases. The simulation thus represents a use case common in chemometrics, with $p > n$ and the ratio of uninformative variables to informative variables varies from 2:1 to 1:1. Data are simulated from a three component model. The dimensionality of the dependent block is kept constant at $q_1 = 3$ informative and $q_2 = 2$ uninformative variables. In each simulation run, sparse twoblock models are estimated with settings $h=3, g=1, \eta=\kappa=0.5$ and a three component PLS2 model is computed. The PLS2 model is estimated using the NIPALS algorithm \citep{WoldNIPALS}. Figure \ref{fig:n_p2_200_MSEB} shows that, as the proportion of informative variables increases, both estimators have increasingly less difficulty to estimate the regression coefficients. However, the new sparse twoblock estimator is capable of estimating the regression coefficients much more accurately (viz. subplot Figure \ref{fig:n_p2_200_MSEB}). Figure \ref{fig:n_p2_200_FPX} illustrates that even in the case where there are twice as many uninformative variables as there are informative ones, only about 2.5\% of uninformative predictors will be included. On the contrary, the number of informative variables left out of the model increases as a higher proportion of such variables is present, with the sparse model leaving out about 10\% of informative variables when $p_1 = p_2$. Subplot \ref{fig:n_p2_200_FY} shows that the model did not include any of the uninformative dependent variables and only sporadically leaves an informative one out. 

The trends plotted in Figure \ref{fig:n_p2_200} generalize to other data dimensionalities. At first, a simulation is presented that allows to investigate a use case that should be less challenging to model: $n > p$, with $n = 1000$ and in the predictor block, $p_1$ was varied from 3 to 20 informative variables and in each simulation, only $p_2 = 2$ uninformative variables were added. Secondly, a simulation is presented, wherein the sample size is fixed at $n=100$ and the dimensionality of the predictor block is increased from 100 to 200 informative variables, while in each run 20 uninformative predictors are being added. The remainder of the settings are the same as in \ref{fig:n_p2_200}. The trends observed in the subplots in \ref{fig:np2_20} are in line with the ones discussed for \ref{fig:n_p2_200}. The same holds true for the trends observed for false positive and false negative rates (not plotted).          

\begin{figure}
\centering
\begin{subfigure}{.4\textwidth} 
\includegraphics[width=\textwidth]{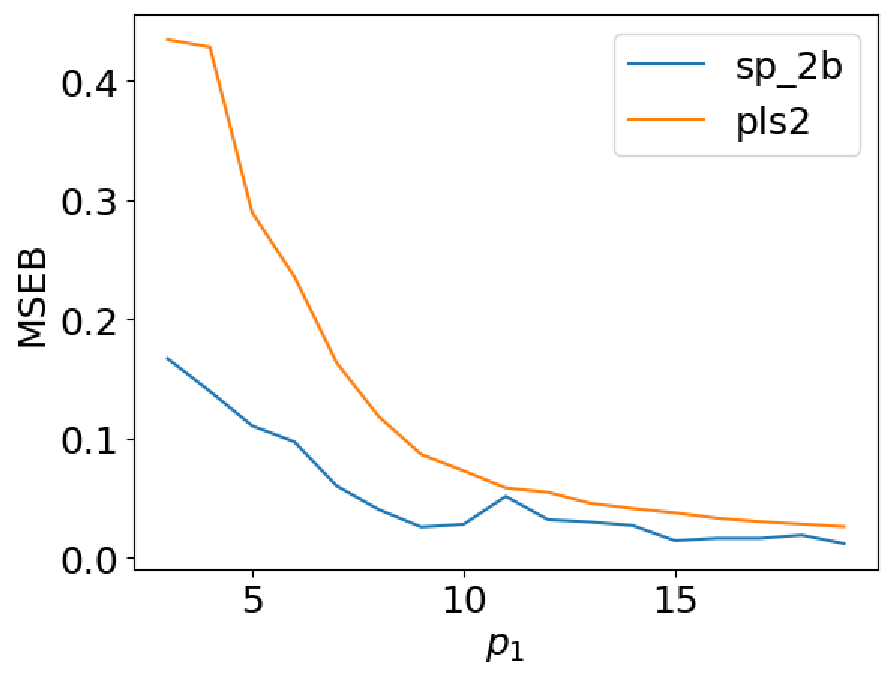}
\caption{$n=1000, p_2=2$}
\label{fig:MSEB_nlp}
\end{subfigure}
\begin{subfigure}{.4\textwidth} 
\includegraphics[width=\textwidth]{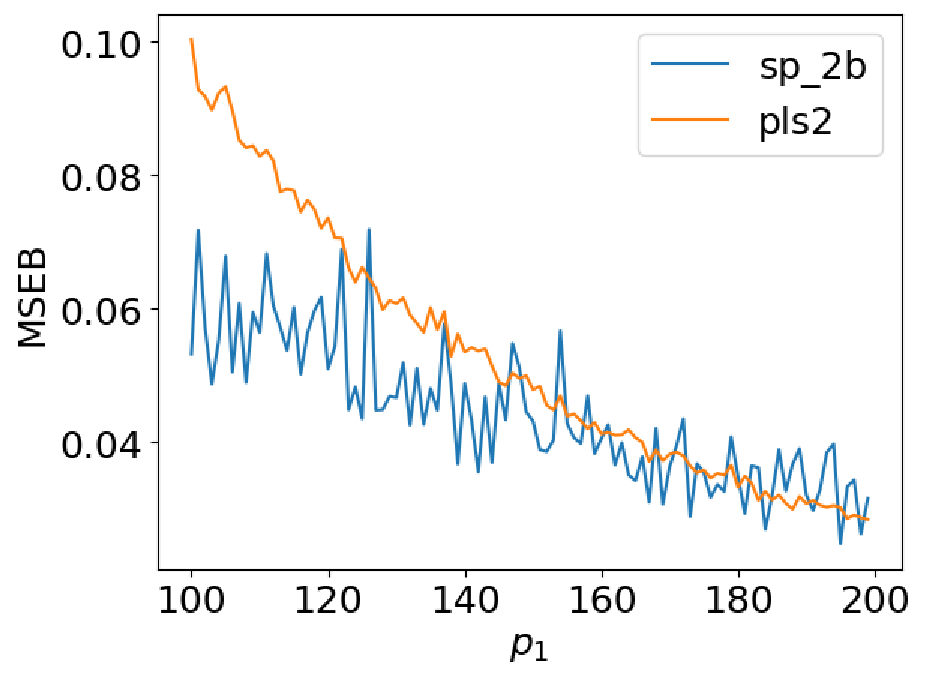}
\caption{$n=100, p_2=20$}
\label{fig:np2_20_MSEB}
\end{subfigure}
\caption{\label{fig:np2_20} Averages over 1000 simulation runs of mean squared error of the regression coefficients (MSEB) as a function of the number of informative predictor variables $p_1$, for two different data dimensionalities. In each case, $q_1=3, q_2=2, g=1, h=3$. }
\end{figure}

Summarizing, the sparse twoblock estimator is suitable to estimate multivariate regression coefficients. It increasingly outperforms PLS2 as the number of uninformative predictor variables increases and it only scantly identifies uninformative variables as informative, or the other way around.  

\section{Examples}

\subsection{Concrete slump data set}

The first example will illustrate that the new algorithm can improve predictive performance, even for data of very moderate dimensions. The data, first reported in \cite{YEH2007474}, consist of seven measurements of compositions pf seven constituents in concrete slumps (cement, slag, fly ash, water,	superplasticizer,	coarse aggregates and	fine aggregates) and three response variables that quantify the slump's physical properties. The original data set consisted of 78 cases. At a later point in time, 25 additional cases were measured. In what follows, the latter will be used as a naturally independent test set. 

Four types of models were cross-validated on the training set: PLS1 models fit to each of the dependent variables individually, a multivariate PLS2 model, dense XY-PLS and finally the Sparse Twoblock algorithm introduced here. At this point, we note that an older dense twoblock PLS method exists \citep{wegelin2000survey}. However, as that method is not sparse and as simulations reported in in \cite{cook2023partial} have shown it to be inferior to XY-PLS, it will not be included in the results discussed here. Tenfold grid search cross-validation was used to determine each model's optimal set of hyperparameters and the optimal model was used to predict responses in the independent test set. Results are reported in Table \ref{tab:concrete}. 

\begin{table}
\begin{center}
\begin{tabular}{|c|c|cccc|}
\hline
Method & Parameters & \shortstack{Slump\\ (cm)} & Flow rate & \shortstack{Compressive\\ Strength} & Average \\
\hline
PLS1 & $h=1,5,3$ & 72.21 & 142.59 & {\bf 6.33} & 73.12 \\
PLS2 & $h=4$ & 61.16 & 176.89 & 6.59 & 81.55 \\
XY-PLS & $g=2, h=5$ & 55.23 & 145.03 & 16.5 & 72.25\\
Sparse Twoblock & \shortstack{$g=3, h=5$,\\ $\eta=.55, \kappa=0.75$} & {\bf 53.21} & {\bf 128.45} & 11.19 & {\bf 64.29}\\
\hline
\end{tabular}
\end{center}
\caption{\label{tab:concrete} Predictive performance on the independent test set set for each of the three dependent variables of the Concrete Slump data set, reported as mean squared error values for the four methods evaluated, as well as the average across all dependent variables.}
\end{table}

The results in Table \ref{tab:concrete} illustrate that the sparse twoblock algorithm can already be very useful for data of limited dimensions. PLS2 struggles most to model all of these variables simultaneously. XY-PLS demonstrates a moderate benefit of the simultaneous dimension reduction in predictor and predictand blocks, whereas the sparse twoblock method introduced here achieves a better performing simultaneous dimension reduction and better overall predictive performance, except for the response variable ``compressive strength.'' However, we note that in the original publication, there was much more emphasis on slump flow than on the two other dependent variables, a variable much better predicted by sparse twoblock then by the other three methods.    

\subsection{NIR Biscuit Dough Data Set}
The NIR biscuit dough data set provides an opportunity for a clean comparison between dense XY-PLS and the new sparse twoblock algorithm, since it is both a frequently cited benchmark data set in chemometrics and was also used as example data set in the seminal publication on XY-PLS \citep{cook2023partial}. The NIR biscuit dough data consist of 72 NIR spectra taken from cookie dough samples at wavelengths in a 2nm increment between 1100 and 2498 nm, on the one hand, and results from wet analytical chemical measurements on the other hand. The objective in the original publication on this data set \citep{osborne1984application} was to build one or more multivariate calibration models that allowed to predict the four reference measurements (fat, flour, sucrose and water) from the spectra, such that the wet chemical measurements could be substituted in online production with less costly and faster measurement of NIR spectra. An occasional spot-check from wet chemical analysis could then confirm the continued validity of the multivariate calibration models. 

In the original publication, the data are split into a training set of 40 samples and a test set of another 32 samples. These data sets were originally recorded at disjoint points in time and not the result of random splitting. Moreover, it was reported that sample 23 in the training set and sample 21 in the test set are outliers, which are typically removed prior to further data processing, as is the case in \cite{cook2023partial} and in the results reported here. 

Using this split into clean training and test data sets, consisting of 39 and 31 cases, respectively, the following models will be compared: univariate PLS (PLS1) regression for each dependent variable separately, multivariate PLS (PLS2) regression based on the NIPALS algorithm \citep{WoldNIPALS}, XY-PLS and sparse twoblock PLS. The latter three methods are all applied to model all four response variables simultaneously. All methods were subjected to a 5-fold cross-validation and the optimal cross-validated parameters are reported along with the predictions. The results are summarized as $R^2$-values for each dependent variable in Table \ref{tab:CookieR2}. 

\begin{table}
\begin{center}
\begin{tabular}{|c|c|cccc|}
\hline
Method & Parameters & Fat & Sucrose & Flour & Water \\
\hline
PLS1 & $h=7,6,6,7$ & {\bf 0.979} & 0.935 & 0.730 & 0.916 \\
PLS2 & $h=6$ & 0.550 & 0.948 & 0.746 & 0.658 \\
XY-PLS & $g=2, h=12$ & 0.947 & 0.904 & 0.838 & 0.897\\
Sparse Twoblock & $g=2, h=9, \eta=.5, \kappa=0$ & 0.930 & {\bf 0.962} & {\bf 0.931} & {\bf 0.948}\\
\hline
\end{tabular}
\end{center}
\caption{\label{tab:CookieR2} Predictive performance on the independent test set set for each of the four dependent variables of the NIR Cookie Dough Data Set, reported as $R^2$ values for the four methods evaluated.}
\end{table}

The results in Table \ref{tab:CookieR2} corroborate the assertion from \cite{cook2023partial} that there is a clear advantage to compressing the responses as opposed to only the predictors. Both methods that accomplish simultaneous dimension reduction, XY-PLS and Sparse Twoblock PLS, clearly outperform PLS2, which struggles much more to maintain predictive power across all four responses. Moreover, the Sparse Twoblock algorithm introduced herein also outperforms univariate PLS for three of the four dependent variables, which may be because it manages to leverage the information contained in the multivariate interplay between the variables. The quality of the predictions are also visualized in Figure \ref{fig:Cookie_Parity}. 

\begin{figure}
\includegraphics[width=\textwidth]{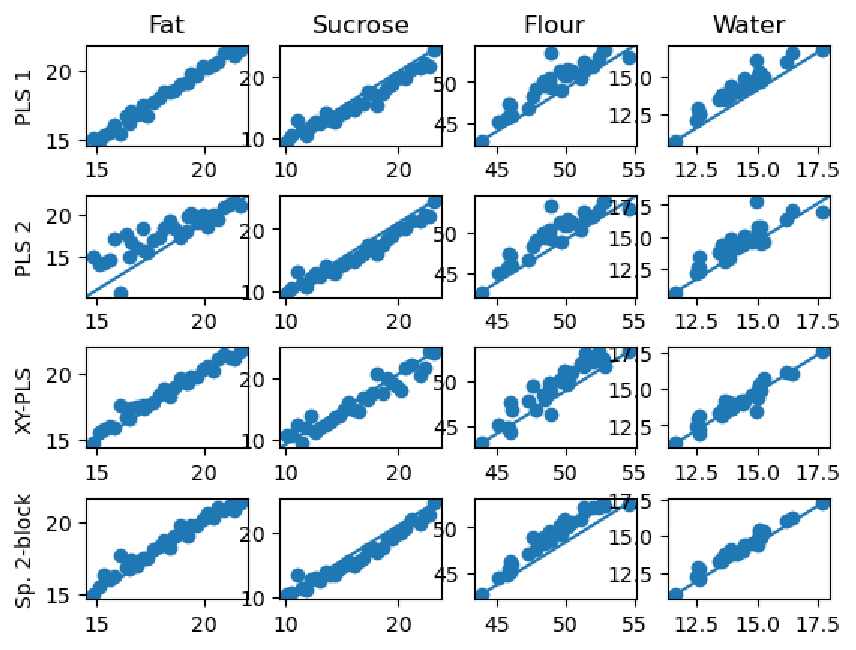}
\caption{\label{fig:Cookie_Parity} Actual versus predicted values for the independent test set in the NIR Biscuit Dough data set, as predicted by PLS1, PLS2, XY-PLS and the novel Sparse Twoblock algorithm.}
\end{figure}

While the Sparse Two-block algorithm yields a slightly lower $R^2$ value for predictions of fat content, from an optical perspective the result is still very close to the one delivered by PLS1. Moreover, the predictions from Sparse Twoblock are also visually less biased for the variables that are more challenging to predict, such as water content.  

The fact that a sparse simultaneous dimension reduction of both blocks outperforms the dense version, is not unexpected either, since \cite{cook2023partial} already report results for XY-PLS based on only 63 {\em central} wavelengths. They decided to trim down the wavelengths used for modeling following an older observation from \cite{li2007partial}, who reported that the wavelengths at the outside of the spectrum contribute little information. 

The introduction of the sparse twoblock algorithm now eliminates the necessity to manually remove wavelengths, as cross-validating the algorithm will accomplish that for the user. One can expect such an algorithmic selection to be as good as expert selection. In fact, in this case, it clearly outperforms: \cite{cook2023partial} report XY-PLS, based on 63 selected wavelengths, to attain a mean squared error of 0.95 (across all four variables). PLS2 attains an MSE of 1.73, whereas the sparse twoblock algorithm introduced herein, attains 0.34. 

As to the wavelengths selected, Sparse Twoblock does trim more than the outskirts of the spectrum. In total, 378 wavelengths were deselected, which amounts to a little over half of all available wavelengths, which is still clearly more abundant than the 64 wavelengths selected by \cite{cook2023partial}. 

\section{Conclusions and outlook}

This paper has introduced a novel algorithm for simultaneous sparse twoblock dimension reduction, that also provides a multivariate regression estimator with intrinsic variable selection capability. The algorithm has been shown to outperform other multivariate regression estimators based on dimension reduction, thereby illustrating that sparse simultaneous dimension reduction can purport an advantage to prediction beyond the potential to yield a highly interpretable dimension reduced space. 

The new method has tuneable sparsity in both predictor and predictand blocks, which allows for variable selection in the {\em dependent} variable space as well. Section \ref{sec:simul} has touched upon identifying uninformative dependent variables, but this aspect was not investigated in the Example section, since in chemometrics, the target of predictive modeling is typically to have models that yield the most accurate predictions for {\em all} of the dependent variables. In chemometrics, a dependent variable can rarely be discarded, if ever. However, that situation is much more frequent in bioinformatics and metabolomics, where applications exist in which the existence of relationships between hundreds of genes and gene expressions have to be analyzed, but no prior knowledge exists as to which variables in each of the blocks are relevant. In such a context, classical \citep{wilms2015sparse} or robust \citep{wilms2016robust} sparse canonical correlation analysis is often applied. A disadvantage of CCA, though,  is that it requires a single number of latent variables to be estimated for both predictors and predictands, even though the dimensionalities of each of these blocks may be very different. The sparse twoblock algorithm introduced here could allow to estimate sparse dimension reduction more efficiently, by allowing for a different number of latent variables in each block. We hope that this and similar ideas will give the method acceptance beyond the illustrative examples touched upon here.

\bibliographystyle{apalike}
\bibliography{bibliography}
\nocite{*}

\end{document}